\documentstyle[graphicx,twoside]{jphysfr}
\makeatletter
\def\theequation@prefix{\thesection.}%
\def\theequation{\theequation@prefix\arabic{equation}}%
\@addtoreset{equation}{section}
\makeatother

\begin{document}
\setcounter{startpage}{1877}
\setcounter{lastpage}{1896}
\setcounter{vol}{4}
\setcounter{num}{11}
\setcounter{tome}{2}

\def\year{1994}
\def\month{11}

\shorttitle{A MODULATED MIRROR FOR ATOMIC INTERFEROMETRY}

\title{A modulated mirror for atomic interferometry}

\author{C. Henkel \inst{1}, A.M. Steane \inst{2}, R. Kaiser \inst{1} 
and J. Dalibard \inst{2}}

\institute{%
\inst{1}
Institute d'Optique Th\'eorique et Appliqu\'ee%
\footnote{Unit\'e de recherche associ\'ee au CNRS.}, 
B\^atiment 503, Centre Scientifique d'Orsay, 91403 Orsay Cedex, France
\\
\inst{2}
Laboratoire Kastler-Brossel
\footnote{Unit\'e de recherche de l'Ecole Normale Sup\'erieure et de l'Universit\'e Pierre et Marie Curie, associ\'ee au CNRS.},
24 rue Lhomond, 75231 Paris Cedex 05, France}

\date{(Received 22 June 1994, 
received and accepted 7 September 1994)%
%
}
\maketitle


\begin{abstract}
A novel atomic beam splitter, using reflection of atoms off an evanescent light wave, is investigated theoretically. The intensity or frequency of the light is modulated in order to create sidebands on the reflected de Broglie wave. The weights and phases of the various sidevands are calculated using three different approaches: the Born approximation, a semiclassical path integral approach, and a numerical solution of the time-dependent Schrödinger equation. We show how this modulated mirror could be used to build practical atomic interferometers.
\end{abstract}

\section{Introduction.}

A number of experimental techniques have been developed to enable the
interference of atoms to be observed, as the present special issue
illustrates. The essential requirement for producing quantum interference is
that a system can pass between two points in its configuration space $\sl
via\rm$ more than one path --- that is, the quantum amplitude for a passage $\sl
via\rm$ either path is non-negligible for a given evolution of the system. The two paths can be visualized as forming a closed loop in configuration-space. If the relative phase between the two quantum amplitudes can vary, then the system can finish in one of two (or more) final states, with probabilities depending on this phase.

The well-known features just mentioned are illustrated by the three main elements of a typical particle interferometer: a first beam splitter, one or more mirrors to bring the two paths back together in  position, and a second beam splitter to close the loop. Diffraction or refraction can also be used to play the role of  a ``mirror'' in bending an otherwise straight path for the interfering particle.

In this paper we describe a new beam splitter for neutral atoms, and also show
how a practical interferometer can be made, having a number of promising
properties such as simplicity, and fairly high (about 8\%) transmission
efficiency into the useful output states. The beam splitter is a vibrating
mirror --- that is a surface which reflects atom incident upon it, and which
moves rapidly to and fro along the normal direction. Such a beam splitter
can form the basis of a number of interferometer designs. An especially
interesting possibility is a very simple interferometer having only a $\sl single\rm$ ``optical element'' --- a horizontal mirror is used, and atoms bounce repeatedly on it. Gravity plays the essential role of bringing the atomic trajectories back to the mirror surface, and the same vibrating mirror is used to separate and recombine the interferometer arms. 

The ability to create a variety of motions of the mirror surface enables one
to manipulate the reflected de Broglie waves in a general manner --- both
delicate adjustments and large shifts of the atomic momentum can be
produced. In the general case, one notes that at normal incidence, the path
length for a wave to travel along the $z$-axis from a position $z$, be reflected
at $z_{\rm m}$ , and then return to $z$, is $2(z-z_{\rm m})$. By varying $z_{\rm m}$ in time, the
variation in path length can be understood as forming an ``optical element''
such as a prism, lens, or phase grating. Here the ``lens'' (or other element) is
extended in $\sl time\rm$ and affects the motion along the mirror normal $z$,
while a conventional lens is extended in $\sl position\rm$ and affects the motion transverse to its axis. Narrow slits and amplitude gratings can be made by switching the mirror reflectivity between one and zero. Our proposal relies on the possibility of vibrating the mirror rapidly (typical vibration frequencies are in the MHz region). This would be difficult for a traditional mirror made of matter, but is easy to achieve for a mirror formed by a light field, since laser beam intensities or frequencies can be modulated rapidly using acousto-optic modulators.

In the following, we first briefly consider beam splitter in general, and the
basic principle of the vibrating mirror (Sect.\ 2). An ideal mirror for atoms
would consist of a sharp potential barrier [1]. In practice, one cannot find
an ideal mirror, and an important type of mirror for atoms is a quasi-resonant
evanescent light wave at the surface of a dielectric [2]. This produces a
potential $V\exp(-2\kappa z)$ having an exponential dependence on atomic
position $z$. We consider atoms bouncing on such a potential, and calculate the
effect of a time-modulation of the amplitude $V(t)$ --- for the exponential
function, such a modulation is equivalent to moving the potential along the
$z$-axis by $z_{\rm m}$ given by

\begin{equation}
V(t){\rm e}^{-2\kappa z} \equiv V(0){\rm e}^{-2\kappa (z-z_{\rm m}(t))} \qquad\mbox{or}\qquad
2\kappa z_{\rm m}(t)=\ln(V(t)/V(0)).
\end{equation}

In sections 3 and 4 we consider two perturbative methods to calculate the probability for an atom to be scattered by the vibrating mirror from one energy eigenstate (or plane wave) to another. The first method uses the Born approximation in first order;  this approximation is valid when the first order scattering probability is low. The second method calculates the phase accumulated by an atom undergoing reflection from a vibrating mirror, using an action integral along the classical path of an atom reflected by a stationary mirror. The range of validity now includes the physically interesting situation where the scattering probability from one eigenstate to another is of the order of 1. In section 5 we consider the case of an initial wave-packet rather than a single plane wave, and compare the approximate analytic results with those of a numerical solution of the Schr{\"o}dinger equation. This enables us to confirm the validity of the various methods. In section 6 we then go on to consider the application of these ideas to make a realistic atom interferometer.

\section{Atom beam splitters.}

Up to now many atomic beam splitters have been suggested, and several
demonstrated. In the following list we mention the beam splitter used in
existing interferometers. These are, to our knowledge, the longitudinal Stark
method [3]; the simple Young's slits arrangement [4, 5]; micro-fabricated
gratings [6]; the Raman pulse technique [7, 8]; the optical Ramsey
interferometer [9, 10]; and the longitudinal Stern-Gerlach interferometer [11]. There has been rapid progress and some very promising results, in term of the minimum detectable phase shift in a given integration time, combined with a large effective area of the interferometer --- both are important since for most experiments the phase change produced by the effects to be measured is proportional to the area of the loop in parameter-space.

Methods to produce larger beam separations have been investigated, notably
adiabatic passage in a ``dark'' state [12, 13]; the magneto-optical beam
splitter [14], and Bragg reflection at crystalline surfaces [15]. The Raman
pulse method has also already been used to produce very large splittings by
the use of many pulses. The vibrating mirror can produce a beam separation
$\Delta p$ of the order of several $\hbar \kappa$ for an incident momentum
$p=100 \hbar \kappa$, as we will show. This allows a useful effective area for
the interferometer without making it too sensitive to misalignments. (The
parameter $\kappa$ in Eq. (1.1) is of the order of the wave vector for light in
resonance with the atomic transition, so $\hbar \kappa$ is approximately equal to the familiar ``recoil'' momentum.)

The basic idea of the vibrating mirror is familiar from optics. To calculate
the effect of a reflection from a mirror whose position $z_{\rm m}$ varies
sinusoidally, $z_{\rm m}(t)=z_0\sin\omega t$, we assume the incident and refected waves can be written
\begin{equation}
\phi_{\rm inc}(z,t)=\exp i(-kz-\Omega t)
\end{equation}
\begin{equation}
\phi_{\rm refl}(z,t)\simeq \exp i(kz-\Omega t-u\sin\omega t+\pi)
\end{equation}
\noindent
The reflected wave here is an approximate solution of the wave equation, valid
when $u\omega \ll \Omega$. We look for a solution having a node on the mirror surface:
\begin{equation}
\phi_{\rm inc}(z_{\rm m}(t),t)+\phi_{\rm refl}(z_{\rm m}(t),t)=0
\end{equation}
\noindent
This implies that
\begin{equation}
u=2kz_0
\end{equation}

The reflected wave has a carrier frequency $\Omega$ plus a frequency modulation imposed by the mirror. It can be decomposed into its component frequencies as follows:
\begin{equation}
\exp i(-\Omega t-u\sin\omega t)=\exp(-i\Omega
t)\sum_{n=-\infty}^{\infty}J_n(u)\exp(-in\omega t).
\end{equation}
\noindent
The weight of a given sideband $\Omega\pm n\omega$ is thus given by $|J_n(2kz_0)|^2$.

Our vibrating mirror for matter waves works along the same general principles. An atom ar-riving with the momentum $p_{\rm i}$ has an energy $\hbar\Omega=p_{\rm i}^2/2M$. After reflection its final momentum $p_{\rm f}$ is given by
\begin{equation}
\frac{p_{\rm f}^2}{2M}=\frac{p_{\rm i}^2}{2M}+n\hbar\omega
\end{equation}
\noindent
where $n$ is a positive or negative integer. For $\hbar\omega\ll p_{\rm i}^2/2M$, the
momentum transfer $\Delta p=p_{\rm f}-p_{\rm i}$ is simply given by
\begin{equation}
\Delta p\simeq n\frac{\hbar\omega M}{p_{\rm i}}=nq
\end{equation}
\noindent
where the elementary momentum spacing $q$ is equal to 
\begin{equation}
q\equiv\frac{\hbar\omega M}{p_{\rm i}}
\end{equation}

The efficiency of the transfer from $p_{\rm i}$ to $p_{\rm f}$, for the case of a vibrating exponential potential, is the subject of the following sections. 

\section{Perturbative calculation in the Born approximation.}

In this section we present a perturbative calculation of the momentum transfer for atoms reflected off a modulated evanescent potential. Since the problem is invariant under a translation parallel to the mirror surface, we only consider the motion along the $z$-axis, defined normal to the surface.

\subsection{The coupling between unperturbed states}
The Hamiltonian is split into two parts:
\begin{center}
$H=H_0+V_1$
\end{center}
\noindent
where $H_0$ is the un-modulated part:

\begin{equation}
H_0=\frac{p^2}{2M}+V_0\exp(-2\kappa z)
\end{equation}
\noindent
This Hamiltonian is responsible for the standard reflection of atoms off the evanescent wave [2]. $V_0$ is the light shift of the ground state of the atom at the prism-vacuum interface located at $z=0$. We consider that the atom adiabatically follows the energy state given by (3.1) without emitting any spontaneous photons. For simplicity, we further do not consider possible atom-surface interactions (van der Waals potential, e.g.), supposing that the atom remains far enough from the surface so that these are negligible compared to the evanescent wave potential in $H$. $V_1$ is the modulated part of the atom-laser interaction [16]:

\begin{equation}
V_1=\epsilon V_0\exp(-2\kappa z)\sin(\omega t)
\end{equation}
\noindent
We will treat the case that the modulation amplitude $\epsilon$ is between 0 and 1.

We evaluate the efficiency of the momentum transfer along $Oz$ by calculating
perturbatively the coupling between eigenstates of the unperturbed Hamiltonian
$H_0$ induced by the time dependent part $V_1$. The eigenstates of $H_0$,
characterized by their asympotic momenta $p>0$, are given by [17] ($P\equiv p/\hbar\kappa$ is the scaled momentum):

\begin{equation}
\Psi_P^0(z)=\sqrt{\frac{2P}{\pi L}\sinh(\pi P)}K_{iP}[w(z)]
\end{equation}
\noindent
where $K_{iP}[w(z)]$ is the Bessel $K$-function of imaginary parameter $iP$, and

\begin{equation}
w(z)=\frac{\sqrt{2MV_0}}{\hbar\kappa}\exp(-\kappa z)
\end{equation}
\noindent
These wavefunctions are the eigenfunctions of the unperturbed Hamiltonian $H_0$, normalized in a box between $z=0$ and $z=L\gg\kappa^{-1}$.

\clearpage

According to Fermi's golden rule, the probability for an atom initially in $\Psi_{\rm i}^0(z)$ (corresponding to an incident momentum $p_{\rm i}$) to make a transition to $\Psi_{\rm f}^0(z)$ (corresponding to a momentum $p_{\rm f}$) is given by:

\begin{equation}
W_{\rm fi}=\frac{1}{\Phi}\frac{\pi}{2\hbar}|\langle\Psi_{\rm f}^0|\epsilon
V_0\exp(-2\kappa z)|\Psi_{\rm i}^0\rangle|^2 \rho(E_{\rm f}=E_{\rm i}\pm\hbar\omega)
\end{equation}
\noindent
where $\rho(E_{\rm f}=E_{\rm i}\pm\hbar\omega)$ is the density of states at the final energy:

\begin{equation}
\rho(E_{\rm f}=E_{\rm i}\pm\hbar\omega)=\frac{dn}{dE}=\frac{ML}{\pi\hbar p_{\rm f}}
\end{equation}
\noindent
and $\Phi$ corresponds to the flux of the incident atomic wave:

\begin{equation}
\Phi=\frac{p_{\rm i}}{2ML}
\end{equation}
\noindent
The coupling term in (3.5) can be obtained analytically [18], yielding:

\begin{equation}
W_{\rm fi}=\frac{\epsilon^2\pi^2}{64}\sinh(\pi P_{\rm i})\sinh(\pi
P_{\rm f})\left\{\frac{(P_{\rm i}+P_{\rm f})(P_{\rm i}-P_{\rm f})}{\sinh(\frac{\pi}{2}(P_{\rm i}+P_{\rm f}))\sinh(\frac{\pi}{2}(P_{\rm i}-P_{\rm f}))}\right\}^2
\end{equation}
\noindent
We now discuss this result in the limits of large and small momenta $P_{\rm i}$, $P_{\rm f}$, respectively.

\subsection{Semiclassical limit}
In the limit $P_{\rm i},P_{\rm f}\gg 1$, we may replace by exponentials the sinh function
in (3.8) whose arguments are of the order of $P_{\rm i}$,$P_{\rm f}$. The transition
probability is then approximated by ($\Delta P \equiv P_{\rm f}-P_{\rm i}$):

\begin{equation}
W_{\rm fi}=\frac{1}{4}\epsilon^2 P_{\rm i}^2 \left(\frac{\frac{\pi}{2}\Delta
    P}{\sinh(\frac{\pi}{2}\Delta P)}\right)^2=\frac{1}{4}\epsilon^2
    P_{\rm i}^2\beta^2(\Delta P)
\end{equation}

The function

\begin{equation}
\beta(x)=\frac{\frac{\pi}{2}x}{\sinh(\frac{\pi}{2}x)}
\end{equation}
\noindent
appearing in (3.9) is equal to unity for $x=0$ and decreases exponentially for
$|x|\gg 1$. It describes the decrease of the efficiency of the momentum transfer if the potential undergoes too many oscillations during the reflection of the atom. Indeed the interaction time of the atom with the evanescent wave is given by
                                                                               
\begin{equation}
\tau=\frac{M}{\kappa p_{\rm i}}
\end{equation}
\noindent
and the momentum transfer can be written

\begin{equation}
\frac{\Delta p}{\hbar\kappa}\approx\omega\tau\equiv Q
\end{equation}
\noindent
where $Q\equiv q/\hbar\kappa$. The rapid decrease of the function $\beta(x)$ means that only a few $\hbar\kappa$ of momentum can be transferred efficiently.

In figure 1, we show $W_{\rm fi}$ at $p_{\rm i}=100\:\hbar\kappa$ for the upper and the
lower sideband, as a function of $Q$. Note that the results (3.8) and (3.9)
predict an asymmetry in the transfer efficiency between the upper and lower
sideband $P_{f\pm}=\sqrt{P_{\rm i}^2\pm 2P_{\rm i}Q}$. This asymmetry is due to the fact
that the monemtum transfer $\Delta P_\pm=P_{f\pm}-P_{\rm i}$ is not the same for the
two sidebands \footnote{The asymmetry (3.13) is due to the non-linear dispersion
  relation $\Omega=\hbar k^2/2M$ of de Broglie-waves. For light waves
  propagating in vacuum in one dimension, $\Omega=ck$, and this asymmetry
  vanishes.}:

\begin{equation}
|\Delta P_+|-|\Delta P_-|\approx -\frac{Q^2}{P_{\rm i}}
\end{equation}
\noindent
Since the transfer efficiency (3.9) depends strongly on the momentum transfer, this difference shows up in the ratio $W_+/W_-$ between the weights of the sidebands:

\begin{equation}
\frac{W_+}{W_-}\approx\exp\left(\pi\frac{Q^2}{P_{\rm i}}\right)=\exp\left(\pi\frac{\hbar\omega^2
  M^2}{p_{\rm i}^3\kappa}\right)
\end{equation}

\begin{figure}
\centerline{\resizebox{!}{58mm}{%
\includegraphics*{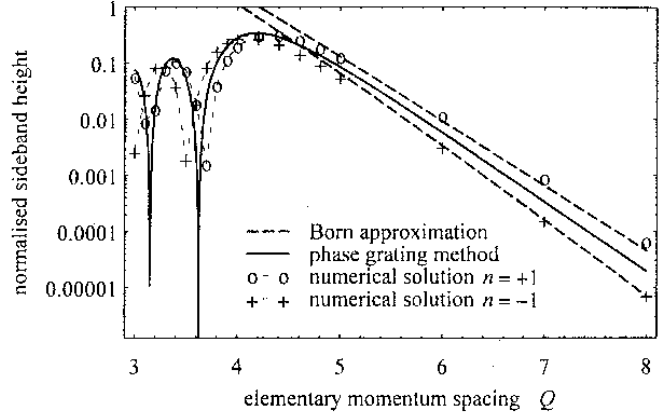}}}
\vspace*{-05mm}
\caption[]{%
Heights of the sidebands for orders $n=\pm1$ as a function of
$Q\equiv\omega\tau$. The incident momentum is $p_{\rm i}=100\:\hbar\kappa$, and the
modulation depth is $\epsilon=1$. The values are normalized to an incident
wave with unit amplitude. Dashed lines: the result of the Born approximation,
$W_{\rm fi}$ (Eq. (3.8)), multiplied by a factor $p_{\rm i}/p_{\rm f}$ in order to normalize
the amplitude of the incident wave. The upper (lower) line corresponds to
$n=+1(-1)$. Solid line: result of the semmiclassical approximation, $|a_1|^2$,
(Eq. (4.20)). Circles: numerical solution (Sect.\ 5) for the normalized momentum
distribution $|\psi(p)/\psi(p_{\rm i})|^2$ with $p^2=p_{\rm i}^2+2M\hbar\omega$ (i.e. sideband $n=+1$). For the sake of clarity, the points have been joined in the region $Q<4.2$ (light dashed line). The parameters of the numerical calculation are given in the text (Sect.\ 5). Crosses: same as before, but now $n=-1$.}
\end{figure}

\subsection{Quantum limit}
In the limit of small momenta $P_{\rm i},P_{\rm f}\ll 1$, one expects the evanescent wave mirror to produce the same resuts as a ``hard'' ideal mirror (Sect.\ 2.), since the wavelength of the incident atom is then longer than the characteristic decay length $1/2\kappa$ of the mirror potential. In this limit expression (3.8) becomes:

\begin{equation}
W_{\rm fi}\approx\frac{1}{4}\epsilon^2 P_{\rm i}P_{\rm f}
\end{equation}
\noindent
This result can be interpreted by noting that the vibrating evanescent wave mirror, in this regime [17], behaves as an ideal mirror moving as 

\begin{equation}
z_{\rm m}(t)=\frac{1}{2\kappa}\ln(1+\epsilon\sin\omega t)
\end{equation}
\noindent
For a small modulation amplitude $\epsilon$, this given a sinusoidal variation
having amplitude $z_0=\epsilon/2\kappa$. Following the discussion of section
2, the weight of the first sideband is then given by $|J_1(u)|^2 \approx u^2/4$
where the argument $u=\epsilon P_{\rm i}$ (using Eq. (2.4) and $k=p/\hbar$). The flux into the sideband $P_{\rm f}$ is then given by (3.15), per unit flux in the incident state.

\section{Semiclassical path integral approach.}

In this section, we present a semi-classical perturbation method which allows
us to derive the atomic wave function after reflection off the modulated
mirror. This method is similar to the one developed for the problem of atomic
wave diffraction by a thin phase grating [19]. It leads to a perturbed wave
function which is phase-shifted with respect to the wave function obtained for
a non-modulated mirror. The phase shift is simply the integral of the
modulated potential along the classical {\it unperturbed\/} trajectory of the bouncing atom. In [19], this method was derived using a Feynman path integral approach which allowed for a detailed study of the approximations involved in the derivation. For the sake of simplicity, we present here an alternative derivation based on a modification of the standard WKB treatment so that it can be applied to a time-modulated potential.

\subsection{Phase shift of the semiclassical wave function}
Consinder the wave function $\Psi^0(z)$ which is an eigenstate of $H_0$ at the energy $E_{\rm i}$. We look for a perturbed wave function of the following form
\begin{equation}
\Psi(z,t)=\Psi^0(z){\rm e}^{-iE_{\rm i}t/\hbar}\exp(iS_1(z,t)/\hbar)
\end{equation}
\noindent
where $S_1(z,t)$ is a small correction introduced by the modulated part of the potential. The time-dependent Schr{\"o}dinger equation gives:
\begin{equation}
\frac{\partial S_1}{\partial
  t}-\left(\frac{i\hbar}{m\Psi^0(z)}\frac{d\Psi^0}{dz}\right)\frac{\partial
  S_1}{\partial z}-\frac{i\hbar}{2M}\frac{\partial ^2S_1}{\partial
  z^2}+\frac{1}{2M}\left(\frac{\partial S_1}{\partial z}\right)^2=-V_1(z,t)
\end{equation}
\noindent
In the semiclassical limit, where the de Broglie wavelenght of the particle is much smaller than the typical length scale $\kappa^{-1}$ of the potential $V_0(z)$, we can approximate $\Psi^0(z)$ in the classically allowed region by the well-known WKB result:
\begin{equation}
\Psi_{WKB}^{0\pm}(z)=\frac{1}{\sqrt{k(z)}}\exp\left(\pm i\int^z k(z')dz'\right)
\end{equation}
\noindent
where the local wave vector $k(z)$ is given by
\begin{equation}
k(z)=\frac{1}{\hbar}\sqrt{2M(E_{\rm i}-V_0(z))}
\end{equation}

Using now this result for $\Psi^0(z)$, we can write
\begin{equation}
-\frac{i\hbar}{M\Psi_{WKB}^{0\pm}(z)}\frac{d\Psi_{WKB}^{0\pm}}{dz}\approx\pm\frac{\hbar
 k(z)}{M}=v(z)
\end{equation}
\noindent
where $v(z)$ is the classical velocity of the particle moving in the unperturbed potential. It can be either negative (before reflection) or positive (after reflection). Note that we have neglected in (4.5) the derivative of the prefactor of (4.3) which would lead to small corrections. The equation of evolution for $S_1$ can now be written
\begin{equation}
\frac{\partial S_1}{\partial t}+v(z)\frac{\partial S_1}{\partial
  z}=-V_1(z,t)+\frac{i\hbar}{2M}\frac{\partial^2S_1}{\partial
  z^2}-\frac{1}{2M}\left(\frac{\partial S_1}{\partial z}\right)^2
\end{equation}
%
\noindent
This equation can be solved formally by the method of characteristics: using the characteristic curve $z_c(t)$ defined by
\begin{equation}
\frac{d}{dt}z_c(t)=v(z_c(t))
\end{equation}
\noindent
the left hand side of (4.6) can be written

\begin{equation}
\left(\frac{\partial}{\partial t}+v(z)\frac{\partial}{\partial
    z}\right)S_1=\frac{d}{dt}S_1(z_c(t),t)
\end{equation}
\noindent
The characteristic curve $z_c(t)$ is the classical trajectory corresponding to a reflection in the unperturbed potential [17]:
\begin{equation}
z_c(t)=\frac{1}{2\kappa}\ln\left[\frac{V_0}{E_{\rm i}}\cosh^2(t-t_0)/\tau\right]
\end{equation}
\noindent
At the ``bouncing time'' $t=t_0$, the atom reaches its classical turning point.

Substituting (4.8) in (4.6), we have the implicit solution

\begin{eqnarray}
S_1(z,t) & = & {} -\int_{-\infty}^t dt'V_1(z'=z_c(t'),t')
\nonumber\\ 
&&
{} +\frac{1}{2M}\int_{-\infty}^tdt'\left(i\hbar\frac{\partial^2
    S_1}{\partial z^2}-\left(\frac{\partial S_1}{\partial
      z}\right)^2\right)(z'=z_c(t'),t')
\end{eqnarray}
\noindent
In the limit $t\rightarrow-\infty$, (4.10) vanishes and (4.1) then reduces to the unperturbed wave function. We are interested here in the case $t-t_0\gg\tau$, when the final time $t$ is in the asymptotic region after the reflection.

We now neglect the second term of the right hand side of (4.10). We will
investigate the validity of this approximation in detail later on, but we can
justify it here in a few words. We are neglecting a second derivative
$\partial^2S_1/\partial z$ and a second order term $(\partial S_1/\partial
z)^2$. The second derivative should have a small contribution in the
semiclassical regime of interest here; its effect is mostly to correct the
classical motion and to change the prefactor entering in (4.3). The second
order term $(\partial S_1/\partial z)^2$ should be small compared to the first
order term entering in (4.2), since we expect $S_1$ itself to be a small
correction to the unperturbed wave function. 

Using the expression (3.2) for $V_1(z',t')$, the integration in (4.10) now yields the phase shift [20]:
\begin{eqnarray}
\frac{S_1(z,t)}{\hbar} & = &
-\frac{\epsilon}{\hbar}\frac{p_{\rm i}^2}{2M}\int_{-\infty}^t\frac{\sin\omega
  t'}{\cosh^2(t'-t_0)/\tau}dt'
\nonumber\\
& = &
-\epsilon P_{\rm i}\beta(\omega\tau)\sin\omega t_0
\end{eqnarray} 
\noindent
Since we have $t-t_0\gg\tau$, this result does not explicitly depend on the upper bound $t$, up to small terms of order $\exp[-2(t-t_0)/\tau]$.

We recover the function $\beta(\omega\tau)=\beta(Q)$ defined in (3.10). The
``bouncing time'' $t_0(z,t)$ of the classical trajectory (4.9) is fixed such that
latter ends at time $t$ at the position $z$: $z_c(t)=z$. Expanding the trajectory (4.9) for $t-t_0\gg\tau$, we find:
\begin{equation}
z=\xi_{\rm eff}+\frac{p_{\rm i}}{M}(t-t_0)
\end{equation}
\clearpage
\noindent
where the ``effective mirror'' position $\xi_{\rm eff}$ equals [17]

\begin{equation}
\xi_{\rm eff}(p_{\rm i})=\frac{1}{2\kappa}\ln\left[\frac{V_0}{4E_{\rm i}}\right]
\end{equation}

We obtain finally
\begin{equation}
\frac{S_1(z,t)}{\hbar}=-u\sin\left(\omega
  t-\frac{q}{\hbar}(z-\xi_{\rm eff})\right)
\end{equation}
\noindent
where the modulation index $u$ is given by:

\begin{equation}
u=\epsilon P_{\rm i} \beta (Q).
\end{equation}

\subsection{The final energy spectrum}
The reflected wave function in the asymptotic region $\kappa z\gg 1$ can now be written

\begin{equation}
\Psi_{\rm fin}(z,t)=C\exp\frac{i}{\hbar}(-E_{\rm i}t+p_{\rm i}z+\hbar\eta+S_1(z,t))
\end{equation}
\noindent
In this expression $C$ is a normalization factor and $\eta=\eta(p_{\rm i})$ is the phase shift of the wave function due to the reflection off the non-modulated evanescent potential. Replacing $S_1(z,t)$ by its expression (4.14), and using (2.5) to expand the result in terms of energy sidebands, we have 
\begin{equation}
\Psi_{\rm fin}(z,t)=C{\rm e}^{i\eta}\sum_{n=-\infty}^{\infty}a_n\exp\frac{i}{\hbar}(-E_nt+p_nz)
\end{equation}
\noindent
where the $n$'th sideband has energy $E_n$ and momentum $p_n(q\equiv\hbar\omega M/p_{\rm i})$:

\begin{eqnarray}
E_n=E_{\rm i}+n\hbar\omega \\
p_n=p_{\rm i}+nq
\end{eqnarray}
\noindent
and its amplitude $a_n$ equals:
\begin{equation}
a_n=J_n(u)\exp(-inq\,\xi_{\rm eff}/\hbar)
\end{equation}
\noindent
Note also that, if the phase of the modulated potential at $t=0$ is shifted by $\phi$, then the phases of the sidebands are shifted accordingly:

\begin{equation}
\omega t\rightarrow\omega t-\phi\qquad\Rightarrow\qquad a_n\rightarrow
a_n{\rm e}^{in\phi}
\end{equation}

As usual in phase modulation problems, there are two limiting regimes for the
result (4.17--4.20), depending on the order of magnitude of the modulation index
$u$. For a low modulation index, $u\ll 1$, the Bessel functions $J_n(u)$ have
magnitude $(u/2)^{|n|}$, and the diffracted spectrum consists essentially of
the carrier $n=0$ and the two first sidebands $n=\pm 1$. In this regime, we recover the result (3.9) of the Born approximation for the first sideband, evaluated for a semiclassical momentum $p_{\rm i}\gg\hbar\kappa$:

\begin{equation}
|a_1|^2=|J_1(u)|^2\simeq\frac{u^2}{4}=\frac{1}{4}\epsilon^2P_{\rm i}^2\beta^2(Q)=W_{\rm fi}
\end{equation}
\noindent
where we have assumed in addition $Q\ll P_{\rm i}$ so that $Q\simeq\Delta P$. Note, however, that the semiclassical approach does not account for the asymmetry of the sideband weights (3.14); this property is related to the approximations involved (see below).

The semiclassical result (4.17) and (4.20) extends the Born result (3.9) into the
region of high modulation index $u\gg 1$, where several lines are present with
an appreciable weight. The most intense lines correspond to $n\sim\pm u$, and
they lead to a velocity change for the atom $\Delta v_{\rm at,max}\simeq
uq/m$. For $\epsilon\ll 1$, where the vibration of $z_{\rm m}(t)$ is harmonic
($2\kappa z_{\rm m}(t)=\epsilon\sin\omega t$), we can relate $\Delta v_{\rm at,max}$ to the ``maximal velocity of the vibrating mirror''
$v_{\rm mir,max}=\epsilon\omega/2\kappa$:

\begin{equation}
\Delta v_{\rm at,max}=2v_{\rm mir,max}\beta (Q)
\end{equation}
\noindent
For a ``hard'' mirror formed with a potential step, one would expect simply
$\Delta v_{\rm at,max}=2v_{\rm mir,max}$. The reduction factor $\beta(Q)$
($\beta(Q)\ll 1$ for $Q>1$), which appears either in the modulation index $u$
(see (4.15)) or in $\Delta v_{\rm at,max}$ (see (4.23)) is due to the ``softness'' of the potential. Indeed, the efficiency of the momentum transfer is proportional to the Fourier transform of the potential ``seen'' by the atom during the reflection (4.11):

\begin{equation}
V_0(z_c(t))=\frac{E_{\rm i}}{\cosh^2(t-t_0)/\tau},
\end{equation}
\noindent
the transform being evaluated at the modulation frequency $\omega$. Since this
Fourier transform has a natural cutoff frequency of the order of $\tau^{-1}$, the transfer efficiency decreases for modulation frequencies larger than this limit, i.e. $Q>1$. The same reduction factor $\beta(Q)$ appears in the classical problem of a particle bouncing on a modulated exponential potential. The equation of motion for the particle is:

\begin{equation}
M\frac{d^2z}{dt^2}=2\kappa V_0{\rm e}^{-2\kappa z}(1+\epsilon\sin\omega t)
\end{equation}
\noindent
The energy change in the reflection can be written:

\begin{equation}
E_{\rm f}-E_{\rm i}=\int_{-\infty}^{\infty}M\frac{dz}{dt}\frac{d^2z}{dt^2}dt=-\epsilon
V_0\omega\int_{-\infty}^{\infty}\cos(\omega t){\rm e}^{-2\kappa z}dt
\end{equation}
\noindent
using an integration by parts. We evaluate this last integral along the atomic trajectory in the non-modulated potential, obtaining

\begin{equation}
\frac{E_{\rm f}-E_{\rm i}}{E_{\rm i}}=-2\epsilon Q\beta(Q)\cos\omega t_0
\end{equation}
\noindent
The maximal atomic velocity change (for $\cos\omega t_0=\pm 1$) is then:

\begin{equation}
\frac{\Delta v_{\rm at,max}}{v_{\rm i}}=\epsilon Q\beta(Q)
\end{equation}
\noindent
or

\begin{equation}
\Delta v_{\rm at,max}=\epsilon\frac{\omega}{\kappa}\beta(Q)
\end{equation}
\noindent
which is identical to (4.23). Note finally that exactly the same dependence on $Q$ appears in the diffraction of atoms by a standing wave at oblique incidence [21, 19].

\subsection{Validity of the semiclassical approach}
The first validity condition for a semiclassical approach requires 

\begin{equation}
P_{\rm i}\gg 1
\end{equation}
\noindent
i.e. an incident de Broglie wavelenght much smaller than the decay lenght of the potential.

A second constraint on the validity of (4.17) results from the use of the method
of characteristics. Since the expression (4.11) for $S_1(z,t)$ has been obtained
by integrating over the classical trajectory of the atoms {\it in the absence\/} of
the modulated potential, we require that this classical trajectory is only
slightly perturbed by the modulation. Otherwise the perturbative expansion
underlying (4.11) would not be possible. This condition is fulfilled in two
cases. One can first take a very small modulated potential ($\epsilon\ll 1$);
the frequency of the modulation can then be chosen freely. The other option is
to take an arbitrarily large modulation factor (up to $\epsilon=1$), but to impose a modulation frequency $\omega$ much greater than the characteristic ``frequency'' of the bouncing process $1/\tau$, where $\tau$ is the reflection time. The modulated potential then induces a fast atomic micromotion, which is superimposed on the slow unperturbed bouncing motion. In the following, we focus on this second option, since it may lead to important transfer of momentum for the atoms. This condition can be written:

\begin{equation}
\omega\tau=Q\gg 1
\end{equation}

In the previous calculation, an additional approximation is involved, which
consists in neglecting in the expression (4.10) for $S_1(z,t)$ the contributions
of the terms $\partial^2S_1/\partial z^2$ and $(\partial S_1/\partial z)^2$. Before going further in a quantitative estimation of the corresponding error, we can point out two consequences of this approximation, which appear clearly in the result (4.17). First, (4.17) does not strictly fulfill the dispersion relation for matter waves $E_n=p_n^2/2M$. Equations (4.18) and (4.19) only constitute a linearized version of this dispersion relation for small $nq$, and this induces a small phase error, that should be compensated by the contribution of these two partial derivatives neglected in (4.10). Secondly the momentum distribution deduced from (4.17) is a symmetric comb centered on $p_{\rm i}$, with sidebands whose weight is proportional to $|a_n|^2=|a_{-n}|^2$. However we know from the Born treatment (Sect.\ 3) that an asymmetry appears in the spectrum, when the parameter $Q^2/P_{\rm i}$ becomes of the order of 1 or larger (see (3.13)). We show now how one can recover this validity condition from a detailed analysis of the contributions of the two partial derivatives mentioned above.

We estimate the magnitude of the second term of the right hand side of (4.10), in order to determine the region of parameter space where it can be neglected with respect to the leading term. As an estimation we take $S_1(z,t)=0$ before the bouncing time ($t<t_0$), and we use for $t>t_0$ the asymptotic expression for $S_1^{(0)}(z,t)$ given in (4.14) \footnote{A more precise evaluation of $S_1$ would require lengthy calculations on this semi-classical framework, due to spurious divergences of the WKB method around the classical turning point.}. We obtain in this way, for $t>t_0$:

\begin{eqnarray}
\frac{\partial S_1}{\partial z}=uq\cos(\omega t_0) \\
\frac{\partial^2S_1}{\partial z^2}=\frac{uq^2}{\hbar}\sin(\omega t_0)
\end{eqnarray}

We now choose a time $t$ such that $t-t_0$ is larger than the reflection time $\tau$ and we evaluate the two contributions that have been neglected in the integral appearing in (4.10):

\begin{eqnarray}
\left|\frac{i}{2M}\int_{-\infty}^t\frac{\partial^2S_1}{\partial
    z'^2}(z',t')dt'\right|\sim u\frac{Q^2}{P_{\rm i}}\frac{t-t_0}{2\tau} \\
\left|\frac{1}{2M\hbar}\int_{-\infty}^t\left(\frac{\partial S_1}{\partial
    z'}\right)^2(z',t')dt'\right|\sim u^2\frac{Q^2}{P_{\rm i}}\frac{t-t_0}{2\tau}
\end{eqnarray}
\noindent
The relative magnitude of these two terms depends on the index of
modulation. Equation (4.34) is the leading term in the regime of low modulation
index ($u\ll 1$), whereas (4.35) is dominant in the high modulation domain
($u\gg 1$).

We note that the magnitude of the two terms (4.34) and (4.35) increases linearly with time, as expected for corrections to the phase of $\Psi_{\rm fin}(z,t)$ whose role is to restore the correct dipersion relation for the de Broglie wave. In practice, we can evaluate these two terms for a time $t$ located well after the classical turning time $t_0$. We take for instance:

\begin{equation}
t-t_0\simeq 4\tau
\end{equation}
\noindent
After the time $t$, the values of the $a_n$ coefficients appearing in the final wave function (4.17) do not change anymore, and we can impose ``by hand'' the correct dispersion relation; in other words, we then replace the wave function (4.17) by

\begin{equation}
\Psi_{\rm fin}'(z,t)=C{\rm e}^{i\eta}\sum_n a_n\exp\frac{i}{\hbar}(-E_nt+p_n'z)
\end{equation}
\noindent
with $p_n'\equiv\sqrt{2ME_n}$.

\begin{figure}[h]
\centerline{%
\resizebox{!}{63mm}{%
\includegraphics*{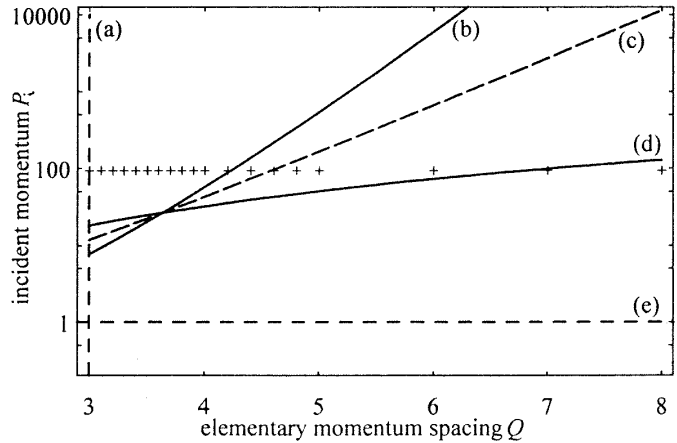}}}
\vspace*{-05mm}
\caption[]{%
Regions of validity of the semiclassical approach, as a function of
$Q$ and $P_{\rm i}=p_{\rm i}/\hbar\kappa$. The modulation depth was taken as
$\epsilon=1$. The semiclassical approach is valid in the region between the
lines (b) and (d). The line (c) ($u=1$) separates the regimes of high (above)
and low (below) modulation index $u$. (a) Line $Q=3$, (Eq. (4.30)) defining the
region for a fast classical micromotion of the atom. (b) Limit $2(\Delta P_{max})^2=P_{\rm i}$ (Eq. (4.39)). (d) Limit $2Q^2=P_{\rm i}$ (Eq. (4.38)). (e) Limiting condition $p_{\rm i}=\hbar\kappa$ for the semiclassical regime. Crosses: parameters corresponding to the numerical solution (Fig.\ 1 and Sect.\ 5).}
\end{figure}

In the regime of low modulation index, the leading correction (4.34) is small compared with the main contribution to $S_1/\hbar$ if: 

\begin{equation}
2Q^2\ll P_{\rm i}
\end{equation}
\noindent
As expected, we recover here the condition required for having a
quasi-symmetric spectrum. 

In the regime of high modulation index, the main contribution to the phase factor $S_1/\hbar$ is very large compared to 1, and we now have to require that the leading phase correction (4.35)
\newpage
\noindent
be small compared to 1. We then obtain:

\begin{equation}
2(\Delta P_{\rm max})^2\ll P_{\rm i}
\end{equation}

\noindent
where $\Delta P_{\rm max}$ represent the maximal appreciable momentum transfer,
measured in units of $\hbar\kappa$:

\begin{equation}
\Delta P_{\rm max}=uQ=\epsilon P_{\rm i}Q\beta(Q)
\end{equation}

\noindent
The validity region delimited by the two equations (4.38) and (4.39), together with (4.30) and (4.31), is plotted in figure 2.

\bigskip

\section{Reflection of wave packets.}

In order to check the approximate results (3.8) and (4.20), we have performed a direct numerical integration of the Schr{\"o}dinger equation, giving the evolution of an atomic wave packet bouncing on the atomic mirror. As in sections 3 and 4, we restricted ourselves to the one-dimensional problem, since the atomic degrees of freedom parallel to the plane of the mirror factorize out. We choose an initial Gaussian wave packet and we integrate the evolution of the atomic wave function in position space using a 4'th order Runge-Kutta algorithm [22].

The position of the initial wave packet is chosen far enough from the mirror
position so that its propagation towards the mirror is initially the same as
for a free particle. For instance, for the evolution shown in figure 3, the
initial wave packet was centered at $\kappa z_{\rm i}=13$, with a standard deviation
$\kappa\delta z_{\rm i}=2$. The initial momentum was $p_{\rm i}=100\:\hbar\kappa$, with a
standard deviation $\delta p_{\rm i}=0.25\:\hbar\kappa$. We have checked that the
final momentum distribution is independent of the value of $\delta p_{\rm i}$,
provided that $\delta p_{\rm i}\ll q$, i.e. in the limit of well-resolved orders.

In our calculation, the particles are confined in a square box whose limits
are $\kappa z_{\rm min}=-25$ and $\kappa z_{\rm max}=25$. The boundary conditions for
the wave function $\psi(z)$ are $\psi(z_{\rm min})=\psi(z_{\rm max})=0$. The
expression of the potential in the $z\ge 0$ domain is the same as the one used
in section 3 (see Eq. (3.1) and (3.2)). In the $z<0$ domain, the potential is
chosen to be null. This form for the potential allows us to study the fraction
of atoms that reach the point $z=0$. In a real experiment these atoms would
actually hit the surface of the dielectric supporting the evanescent light
wave. They would either stick to the dielectric, or be reemitted with a
thermal velocity; in any case they would be lost for the subsequent use of the
bouncing atomic beam. For the values of the parameters used in the following
examples, we have checked that this fraction of atoms is always negligible,
even for a $100\%$ modulated potential ($\epsilon=1$) \footnote{For the practical design of an experiment, one should include, for safety, the van der Waals attractive potential in order to estimate more precisely the fraction of sticking atoms.}.

After the reflection, i.e. when the reflected wave is in a region where the
influence of the potential is negligible, we calculate the momentum
distribution $|\bar{\psi}(p)|^2$, where $\bar{\psi}(p)$ is the Fourier
transform of $\psi(z)$. Two example momentum distributions are shown in figure
4. We have checked that the result for $|\bar{\psi}(p)|^2$ does not depend on
the time at which the Fourier transform is taken, as expected from the simple
evalotion of $\bar{\psi}(p)$ after reflection:
$\bar{\psi}(p,t_2)=\bar{\psi}(p,t_1)\exp(ip^2(t_2-t_1)/2M\hbar)$. On the other
hand, the shape of the position distribution $|\psi(z)|^2$ changes long after
the reflection. For instance the oscillations appearing in figure 3, which
result from the superposition of wave packets with momenta $p_{\rm i}$, $p_{\rm i}\pm q$, will eventually disappear when the three wave packets become separated from each other because of their different group velocities.

\newpage

\begin{figure}
\centerline{%
\resizebox{!}{155mm}{%
\includegraphics*{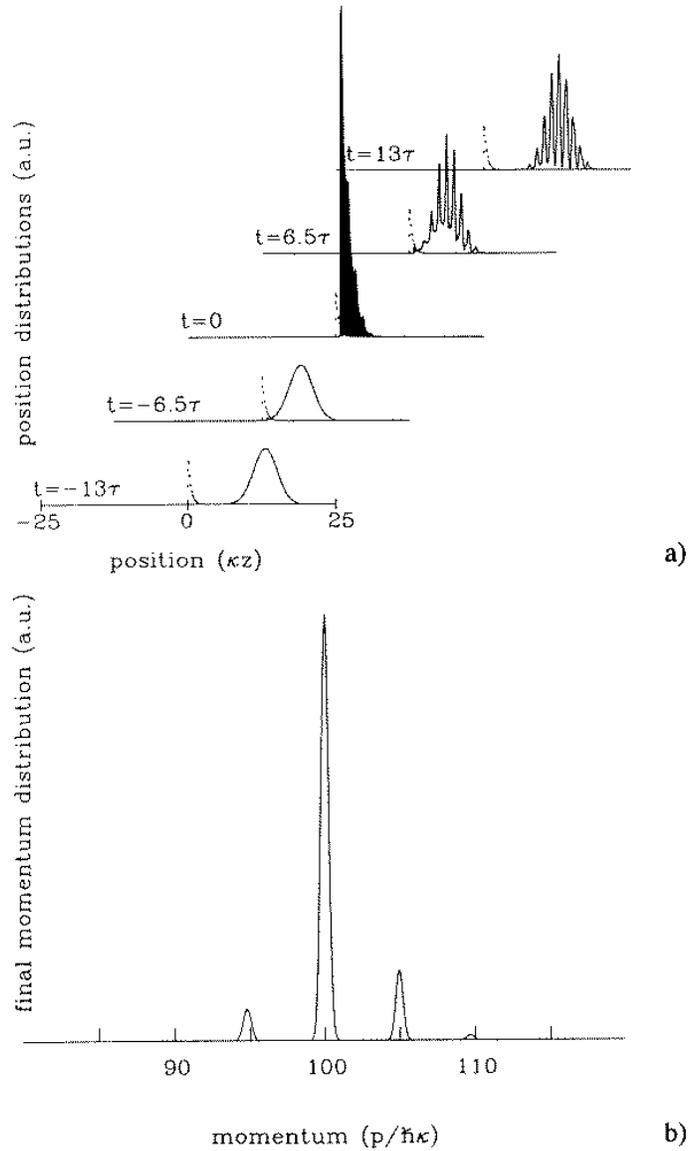}}}
\vspace*{-5mm}
\caption[]{%
a) Reflection of a wave packet at the modulated mirror. The position
distribution at $t=0$ has been cut for the sake of clarity, its actual maximum
height is about 15 times the height of the incident wave packet. The
parameters used are: initial position $\kappa z_{\rm i}=13$ with standard deviation
$\delta\kappa z_{\rm i}=2$, initial momentum $p_{\rm i}=100\:\hbar\kappa$ with standard
deviation $\delta p_{\rm i}=0.25\:\hbar\kappa$, modulation frequency $\omega=5/\tau$ so that $q=5\:\hbar\kappa$, modulation depth $\epsilon=1$. b) Final momentum distribution.
}
\end{figure}

We now compare the predictions of this numerical treatment with those of the approach presented in section 4. To this purpose, we consider again the initial Gaussian wave packet:

\begin{equation}
\bar{\psi}(p,0)=\exp\left(-\frac{(p+p_{\rm i})^2}{4\delta
    p_{\rm i}^2}\right)\exp\left(-\frac{i}{\hbar}pz_{\rm i}\right).
\end{equation}

\begin{figure}
\centerline{%
\resizebox{!}{151mm}{%
\includegraphics*{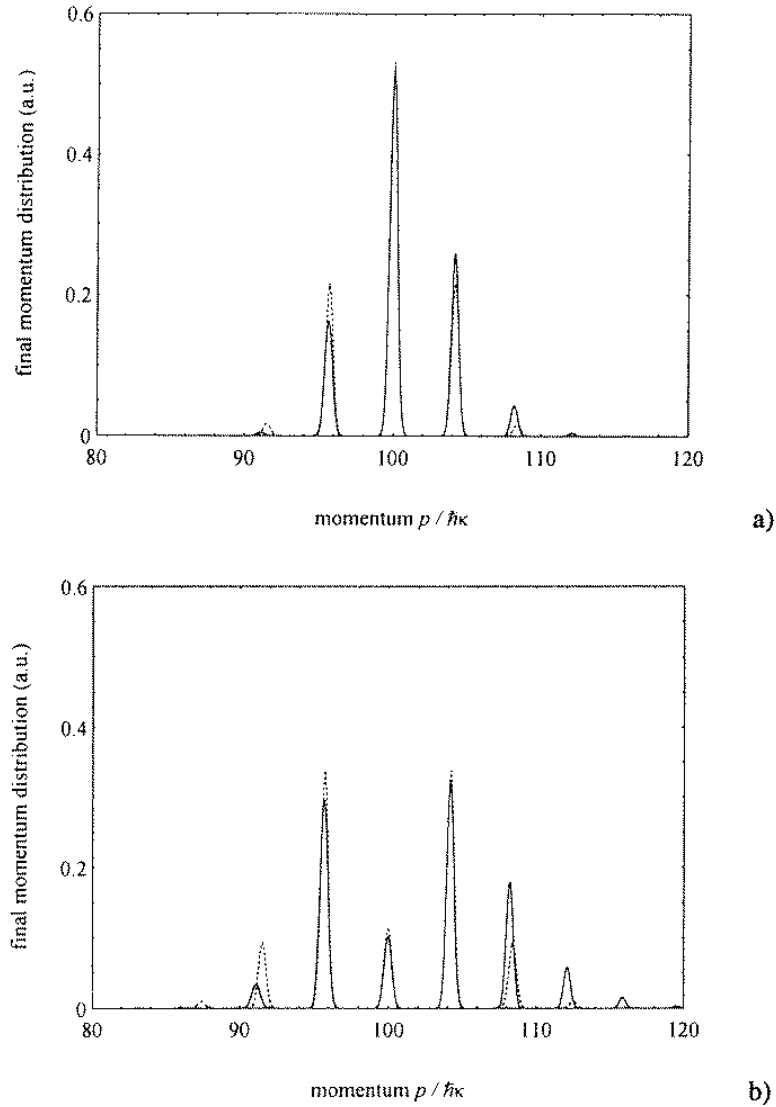}}}
\vspace*{-05mm}
\caption[]{%
Final momentum distributions, given by the numerical solution (solid
line) and the semiclassical approach (dashed line). The parameters used are:
initial momentum $p_{\rm i}=100\:\hbar\kappa$, momentum width $\delta
p_{\rm i}=0.25\:\hbar\kappa$ (standard deviation), modulation frequency
$\omega=4.2/\tau$ (momentum transfer $\Delta p=4.2\:\hbar\kappa$). The situation
adopted in (a) (modulation depth $\epsilon=0.6$) maximizes the product of the
carrier and first sideband; that in (b)($\epsilon=1$) maximizes the product of
the sidebands $n=\pm 1$.}
\end{figure}

\noindent
Using the approximate results of section 4, the final wave packet is given by

\begin{equation}
\bar{\psi}(p,t)=\sum_{n}a_n{\rm e}^{i\eta}\exp\left(-\frac{(p-nq-p_{\rm i})^2}{4\delta
    p_{\rm i}^2}\right)\exp\left(-\frac{i}{\hbar}pz_{\rm i}-\frac{i}{\hbar}\frac{p^2t}{2M}\right)
\end{equation}
\noindent
where the $a_n$ coefficients, given in (4.20), must be evaluated at $p-nq$. The
results of the two approaches are shown in figure 4, the solid line for the
numerical solution and the dotted line for the semiclassical
approximation. The initial wave packet is the same as above. The momentum
transfer is $q=4.2\:\hbar\kappa$ and the modulation amplitude $\epsilon$ equals
0.6 (Fig.\ 4a) and 1 (Fig.\ 4b). We see that the agreement between the
predictions of the two methods is good, although not perfect. In particular,
the numerical treatment shows an asymmetry between the heights of the two
sidebands $|\bar{\psi}(p_{\rm i}\pm q)|^2$, while the approximate treatment predicts equal sideband weights, because of the relation $|J_n(x)|=|J_{-n}(x)|$ (see discussion in Sect.\ 4). 

A more systematic comparison of the predictions of this numerical approach
with the results derived with the Born approximation and with the
semiclassical approach is presented in figure 1. We have determined, for the
same initial wave packet as above, and for several values for $q$, the heights
of the sidebands $n=\pm 1$. We see that the agreement between the three methods, in their expected range of validity, is quite satisfactory.

\section{An atom interferometer.}

The vibrating mirror can be used in a number of ways to make an interferometer. In what follows we will consider the case of atoms normally incident on the mirror surface, and making three or more bounces on it. In other words, we rely on the possibility of having a source of slow atoms released above the mirror [23, 24]. With a fast beam of atom, either which could not be reflected at normal incidence, or for which the time to perform repeated bounces is too long, one would need several mirrors --- used at grazing incidence if necessary.

Figure 5 illustrates an interferometer based on three consecutive bounces on a vibrating mirror. This is similar in conception to an interferometer constructed from three diffraction gratings [6, 25]. For an incident energy $E_{\rm i}$, we consider two output channels with energies equal to $E_{\rm i}$ and $E_{\rm i}+\hbar\omega$. Each of these two channels can be reached via two paths, which are shown by full lines in figure 5; the other paths, shown dashed in figure 5, do not contribute since we assume that the mirror is ``turned off'' when these paths hit it.

\begin{figure}[h]
\centerline{%
\resizebox{!}{54mm}{%
\includegraphics*{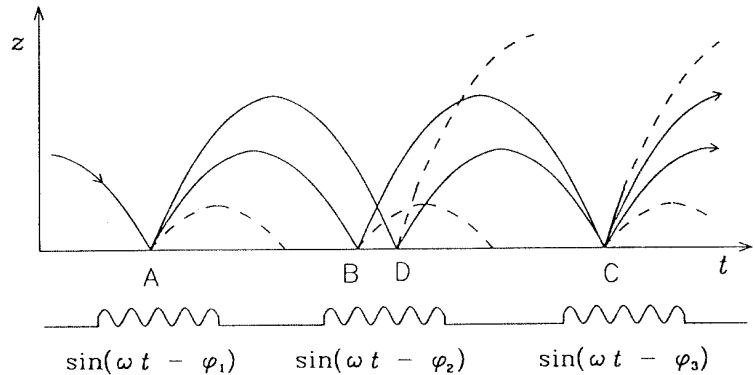}}}
\vspace*{-04mm}
\caption[]{%
Space-time diagram showing a monochromatic  interferometer using three bounces on a vibrating mirror. The paths shown lie at local minima of the classical action. At each reflection, paths corresponding to sidebands of order higher than 1 are omitted for clarity, and paths which return to the mirror when the latter is non-reflecting are shown dashed.
}
\end{figure}

The probability amplitude to exit in a given channel is calculated using the propagators for the region where the atom is in free fall, and the phase shifts corresponding to the interaction with the mirror. The free fall propagator can be evaluated using the integral of the action along the relevant classical paths. We write these in the form $\exp(i\alpha_{AB})$ for the various paths as shown in figure 5. We write $\phi_m$ the phase of the mirror vibration at the $m^{\rm th}$ bounce. This phase is dictated purely by the origin in time of the sinusoidal modulation of the mirror for that bounce. Using the expressions (4.20) and (4.21) for the coefficients $a_n$ giving the amplitude for transmission into the $n$'th sideband, we now get the two probability amplitudes for transmission into the channels:
\begin{eqnarray}
{\cal A}(E_{\rm i}) & = &
a_0^{(1)}a_1^{(2)}a_{-1}^{(3)}{\rm e}^{i(\phi_2-\phi_3)}{\rm e}^{i(\alpha_{AB}+\alpha_{BC})}
\nonumber\\
&&
{}+a_1^{(1)}a_{-1}^{(2)}a_0^{(3)}{\rm e}^{i(\phi_1-\phi_2)}{\rm e}^{i(\alpha_{AD}+\alpha_{DC})}
\\
{\cal A}(E_{\rm i}+\hbar\omega) & = &
a_0^{(1)}a_1^{(2)}a_0^{(3)}{\rm e}^{i\phi_2}{\rm e}^{i(\alpha_{AB}+\alpha_{BC})}
\nonumber\\
&&
{}+a_1^{(1)}a_{-1}^{(2)}a_1^{(3)}{\rm e}^{i(\phi_1-\phi_2+\phi_3)}{\rm e}^{i(\alpha_{AD}+\alpha_{DC})}
\end{eqnarray}
\noindent
where the superscript ($m$) indicates the bounce number. For simplicity we have
omitted here the contribution of the phases $\eta(p)$ which appear in (4.17). This contribution is the same for the two paths and it cancels out in the final interference pattern.

By choosing a symmetric geometry, we impose the conditions
$\alpha_{AB}=\alpha_{DC}$ and $\alpha_{AD}=\alpha_{BC}$ so that the final
interferometric phase is given simply by the reflections; it can be written
$\theta=\phi_1-2\phi_2+\phi_3$. Note that the other phases associated with the
reflections, $q\xi_{\rm eff}$ and, as already noted, $\eta(p)$, exactly cancel out
because of the symmetry. In other words all the phases which depend on the
incident atomic momentum disappear, and the interferometer will produce
high-contrast fringes even when illuminated by a beam of atoms having a broad
momentum distribution. The ``tradional'' three-grating interferometer has the
same property. Note also that, for any phase grating including the present
one, the $a_n$ coefficients satisfy $a_0^*a_{-1}=-a_0a_1^*$. This ensures that
the total probability for ending in one channel or the other, $|{\cal A}(E_{\rm i})|^2+|{\cal A}(E_{\rm i}+\hbar\omega)|^2$, is independent of the interferometric phase $\theta$.

The fringe amplitude for this interferometer can be written

\begin{equation}
{\cal F}=2{\rm Re}
\left[
(a_0a_1^*)^{(1)}(a_1a_{-1}^*)^{(2)}(a_{-1}a_0^*)^{(3)}{\rm e}^{i\theta}
\right]
\end{equation}
\noindent
This quantity would give ${\cal F}=\cos\theta$ for an ideal interferometer such as a Mach Zehnder. Using the semiclassical approximation (4.20), and denoting $u_m$ the index of modulation for the $m$'th bounce, one has
\begin{equation}
{\cal F}=2(J_0(u_1)J_1(u_1))(J_1(u_2))^2(J_0(u_3)J_1(u_3))\cos\theta
\end{equation}
\noindent
This fringe amplitude $\cal F$ is optimised for $u_1=u_3=1.08$ and $u_2=1.84$,
and we get $|{\cal F}|_{\rm max}=0.078$. To maximize in these conditions the
separation of the interferometer arms we choose $\epsilon_2=1$ at the middle
bounce, and (4.15) then gives $\epsilon_1=\epsilon_3\simeq1.08/1.84\simeq0.6$
for the first and third bounces. For example, if the initial momentum is
$p_{\rm i}=100\:\hbar\kappa$, (4.15) leads to $\Delta p=4.2\:\hbar\kappa$ for these values
of the $\epsilon$'s and $u$'s. This gives a good indication of the optimum
case: the numerical solution of the Schr\"odinger equation gives the maximum
$|{\cal F}|_{\rm max}\simeq0.081$ at $\Delta p\simeq 4.2\:\hbar\kappa$, for the same value of $p_{\rm i}$ (see Figs. 4a and 4b). To optimise instead the fringe contrast, one would use the case $|a_0|=|a_1|$ for the first and third bounces, producing $100\%$ contrast, with a slightly reduced fringe amplitude.

As an example, consider a cesium atom normally incident on an evanescent wave
near-resonant with the atomic transition $6S_{1/2}\rightarrow
6P_{3/2}$. Taking $\kappa\simeq 2\pi/852$ nm, the momentum $100\:\hbar\kappa$
corresponds to a velocity of 35 cm/s, and a bounce height of 6 mm, which is
readily realisable in practice. The momentum change $4.2\:\hbar\kappa$ is
produced with $\omega=2\pi\times 1.7$\hspace*{1ex}MHz. The two interferometer arms are separated in distance by about 0.5 mm, when both are near their maximum height, and in time by about 3 ms (the difference in arrival times at the second bounce). Thus macroscopic path separation can be obtained.

\begin{figure}
\centerline{%
\resizebox{!}{51mm}{%
\includegraphics*{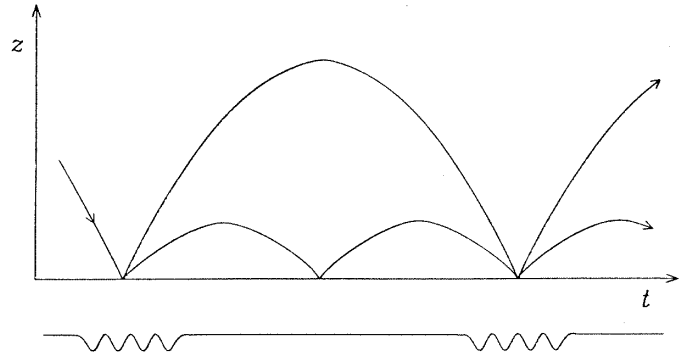}}}
\vspace*{-04mm}
\caption[]{%
Interferometer design in which the interference phase is sensitive
to the acceleration due to gravity. Since the vibrating mirror can only
transfer a few $\hbar\kappa$ of momentum, a practical design would involve several bounces before the arms are recombined, as described in the text.}
\end{figure}

The interference phase will be destroyed not only by mechanical fluctuations
of the dielectric surface supporting the evanescent wave, but also by
intensity fluctuations of the evanescent wave itself. To estimate the effect
of the latter, we note that if the intensity fluctuates by a fraction
$\epsilon$, then the effective mirror position (Eq. (4.13)) moves by
$\epsilon/2\kappa$. The change produced in the interferometer phase is then
$\epsilon p/\hbar\kappa$ radians; this can be understood simply as a higher
sensitivity to fluctuations when the de Broglie wavelenght is small, or,
equivalently, when the interferometer area is large. One must eliminate not
only fluctuations in time but also in the transverse spatial profile of the
laser beam forming the evanescent wave ``mirror''. To make a flat mirror, one
could use a Gaussian laser beam of waist $w$ internally reflected from a concave
glass surface of radius of curvature R: when $w=\sqrt{2R/\kappa}$ the glass
surface curvature compensates the Gaussian fall-off of the laser beam
intensity profile, and the mirror presents a flat reflecting surface for the atoms. A transverse magnetic quadrupole field could help to confine the atom without perturbing the vertical motions of the interferometer arms.

The most obvious first use of the vibrating mirror interferometer is simply to
investigate the preservation of coherence during a reflection off the
evanescent wave mirror --- this could be done by introducing more and more
bounces between the splitting and recombining of the two interferometer arms. A
second use is as a probe of the acceleration due to gravity. For this, one
would not use the symmetric arrangement which we have considered so far, but
instead one allows the two arms to bounce many times until they are recombined
``automatically'' after $m$ (and $m+1$) bounces, where $mp'=(m+1)p$, with $p'^2=p^2+2M\hbar\omega$ (Fig.\ 6). The mirror is vibrated only for the initial and final bounces; in between it is stationary. For this case, the interferometer phase is dominated by the contributions from the free flights between bounces. One finds that while such an interferometer provides a sensitive probe of the acceleration due to gravity, it is also highly chromatic, producing fringes only for a very small class of incident momenta centred around $p$.

\section*{Conclusion.}

To summarize, we have presented here the principles of a vibrating mirror for 
atoms, which constitutes a novel beam splitter for atomic de Broglie waves. We 
have investigated two
analytical theoretical approaches for this new scheme, in order to determine 
the amplitudes and the phases of the outgoing atomic waves. The results are in
excellent agreement with a numerical approach to the problem.

The typical momentum transferred by a vibrating mirror formed with an evanescent light wave is a few photon momenta.\ Although 
this is not as large as the transfers obtained by some\\
other devices, we believe that it should provide, because of its conceptual simplicity, a convenient tool for atom optics and interferometry. Indeed, unlike other atom optics components such as micro-fabricated gratings, it is quite easy to change rapidly (on the microsecond scale) the modulation factor of the light wave forming the mirror. This gives in return a direct control upon the phases and intensities of the diffracted de Broglie waves, and this allows one to conceive simple and useful atomic interferometric devices, such as the ones shown in figures 5 and 6.

\section*{Acknowledgement.}
J.D. thanks C. Cohen-Tannoudji, Y. Castin and P. Storey for useful discussions 
concerning the general problem of modulated potentials in Atom Optics. R.K. 
and C.H. are indebted to J.-Y. Coutois, C. Westbrook and A. Aspect for 
stimulating comments. This work is par\-tially supported by CNRS, DRET, 
Coll\`ege de France, and EEC. A.M.S. is financed by the Commission of European 
Communities through a Community trainig project.


\begin{thebibliography}{99}
\bibitem{[16]} The decay constant $\kappa$ of the modulated potential (3.2) 
may be  different from the one of the static potential. In this work, however, 
we restrict ourselves to identical $\kappa$'s.
\end{thebibliography}
\end{document}